**Analog of the Intertype Superconductivity in Nanostructured Materials**

D.M. Gokhfeld[1,2], S.I. Popkov[1,2], A.A. Bykov[3*]

[1]Kirensky Institute of Physics, Federal Research Center KSC SB RAS, Krasnoyarsk, 660036, Russia

[2]Siberian Federal University, Krasnoyarsk, 660041, Russia

[3]NRC «Kurchatov Institute» - PNPI, 188300, Gatchina, Russia

*Email: redi87@bk.ru

**Abstract**

Magnetization hysteresis loops of tin samples with an inverted opal structure are presented. The sample formed by tin particles with the size of 70 and 128 nm is found to be a type-I superconductor. The tin sample formed by 80 and 42 nm particles demonstrates an analog of intertype superconductivity: features of both type-I and II superconductors are observed on the magnetization isothermal curves. A behavior of the irreversible and reversible magnetizations supports coexistence of type-I and II superconducting nanoparticles in this sample.

**Introduction**

To be superconducting, a material should possess the strong diamagnetic response as well as the zero resistance [1]. While the external magnetic field $H$ is smaller than the thermodynamic critical field $H_c$, supercurrents circulate and the magnetic flux is expelled from type-I superconductors. Type-II superconductors can contain the magnetic flux lines, Abrikosov vortices, thereby samples maintain the superconducting state at higher magnetic fields until the upper critical field $H_{c2}$. Abrikosov vortices emerge when the Ginzburg-Landau parameter $\kappa = \lambda/\xi$ is higher than $1/\sqrt{2}$, here $\lambda$ is the magnetic penetration depth and $\xi$ is the coherent length of the superconductor. Magnetization curves of type-I and II superconductors are schematically shown in Figure 1.

Abrikosov vortices repel usually one from other. In contrast, the coexistence of vortex attraction and repulsion is possible when the $\kappa$ value is about $1/\sqrt{2}$ [2–4]. For such an intertype state (I+II), the field dependence of magnetization $M(H)$ looks like a mixture of magnetizations of type-I and II superconductors [3]. This kind of the magnetization curve is also shown in Fig. 1. A similar attractive-repulsive interaction of vortices was predicted for a two-band superconductor $MgB_2$ [5–7]. Also, it was shown [8] the intertype state is realized by decreasing the cross-sectional area of type-I superconductor nanowires. Arrays of nanowires with various sizes open up perspectives for composite superconducting materials with variable magnetic properties [8].

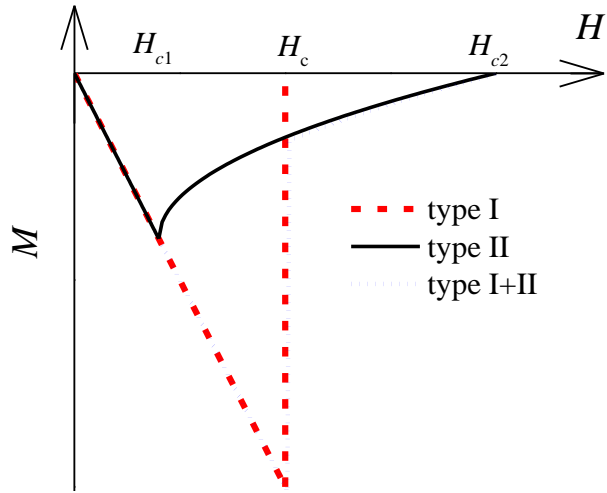

Fig. 1. Schematic magnetization curves for type-I, type-II and intertype superconductors.

Recently, methods have been developed to produce nanostructured superconductors with a structural element size of ~ 10-100 nm [9–15]. Values of $\xi$, $\lambda$ and $\kappa$ differ for nanostructured and bulk superconductors [10,11,16]. Thus, tin particles with the size of 120 nm are the type-I superconductor with $\mu_0 H_c(0) \approx 0.03$ T [13]. This critical field is the same as for bulk tin [17]. The smaller tin particles are the type-II superconductor: the upper critical field $H_{c2}$ reaches 0.15 T and 5.4 T for particles with the size of 58 nm and 8 nm correspondingly [10].

Nanostructured superconductors can be created by pumping a molten metal into an opal matrix. Inverted opal structures (IOS) are resulted, in which a superconducting metal fills the space between closely packed $SiO_2$ spheres. Superconducting properties of IOS were earlier investigated for gallium [14], indium [12] and lead [9] based structures. The investigated samples are found to be type-II superconductors. In presented paper, the tin based IOS samples are investigated. The sizes of structure elements of the samples are larger than ones in work [10], but smaller than ones in [13].

**Materials and Methods**

Opal-like matrices were obtained by sedimentation of the $SiO_2$ spheres and infiltrated with Sn under a pressure. Structural and magnetic measurements were performed for IOS with the $SiO_2$ sphere diameter of 300 nm (the sample S1) and 190 nm (the sample S2). Magnetization curves $M(H)$ were measured by using QD PPMS-6000 magnetometer.

The Sn based IOS samples are three-dimensional networks of connected octahedral and tetrahedral Sn particles [18]. As scanning electron microscopy and small-angle synchrotron scattering revealed, the sizes of octahedral and tetrahedral particles equal 128 and 70 nm for S1 and 80 and 42 nm for S2. The detailed structural and magnetic results will be published in other work [19]. A schematic view of the realized inverted opal structures is shown on Fig 2. The octahedral and tetrahedral particles marked at the primitive cell at Fig. 2 by color.

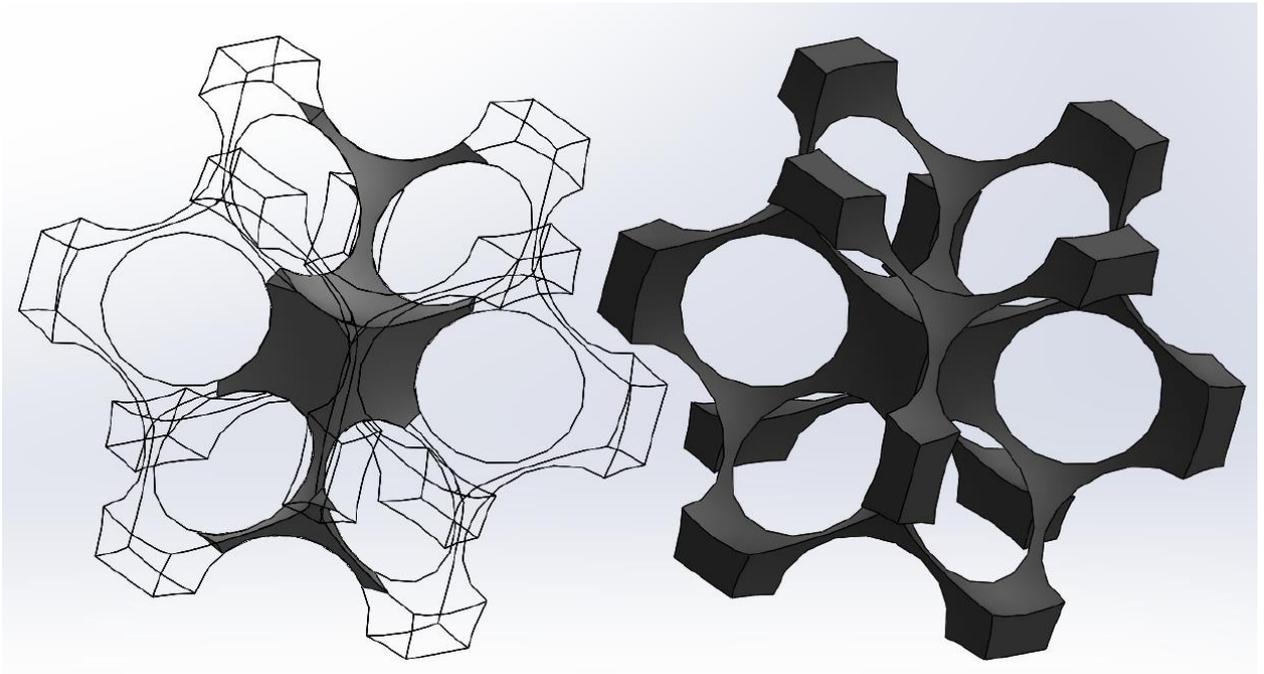

Fig. 2. The scheme of the inverted opal structure.

**Results and discussion**

Magnetization isotherms for the samples at $T = 2$ K are shown in Fig. 3. Characteristic fields of observed features are marked: $H_m$ is the field of a maximal diamagnetic response, $H_{irr}$ is the irreversibility field, where the forward magnetization branch coincides with the reversal branch, $H_c$ is the critical field of the type-I superconductor, $H_{c2}$ is the upper critical fields of the type-II superconductor.

The magnetization curve of S1 has hysteresis in field range from $-H_{irr}$ to $H_{irr}$. When the magnetic field decreases, the magnetization reversal branch has a kink at $\mu_0 H \approx 0.015$ T. The value of $H_c$ is equal to the critical field of the bulk tin and consequently the sample S1 is the type-I superconductor. Type-I superconductors usually have only an equilibrium magnetization, while the hysteresis feature corresponds to a nonequilibrium magnetization. The observed nonequilibrium magnetization of S1 is due to the magnetic flux trapping by superconducting contours [20].

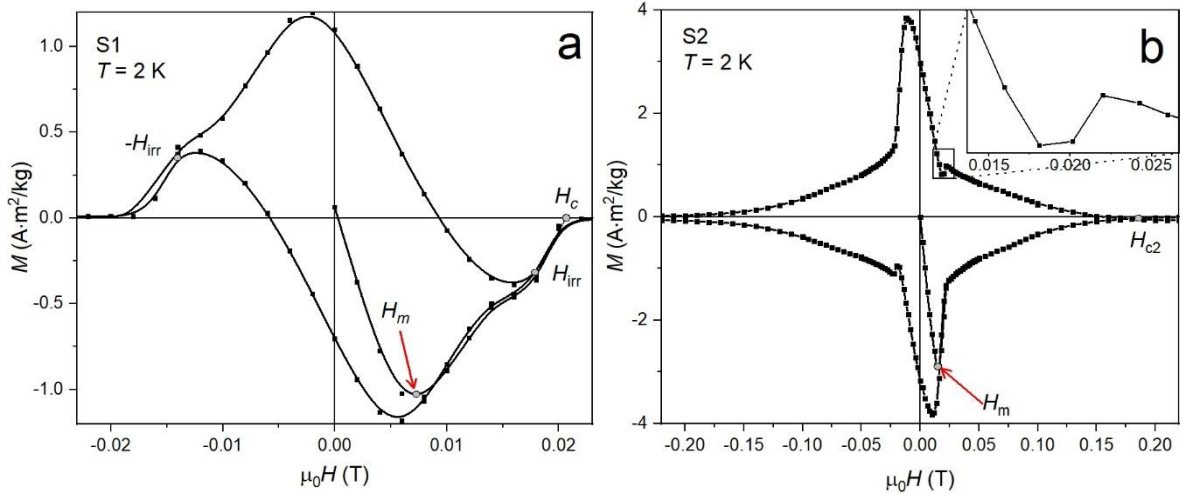

Fig. 3. The magnetization hysteresis loops of the samples S1 (a) and S2 (b).

The $M(H)$ curve of S2 has hysteresis and the high value of the upper critical field, $\mu_0 H_{c2} \approx 0.2$ T, which is much higher than $H_c$ of the bulk tin. Thus, the type-II superconductivity is realized in S2. In contrast to usual magnetization curves of type-II superconductors (Figure 1), the virgin magnetization of S2 has two regions with different slopes: the diamagnetic signal grows at $0 < H < H_m$ and decreases at $H > H_m$ in the similar way as for S1, then the diamagnetic response decreases slowly until $H = H_{c2}$. The reversal magnetization branch also demonstrates two regions: a smooth region at $H_{c2} > H > H_c$ is replaced by a sharp slope at $H_c > H > 0$. The regions with different slopes are separated by a kink at $\mu_0 H \approx 0.18$ T (see insert of Figure 3b). Such magnetization curves are assumed to be an attribute of the intertype superconductivity [3,5]. Recently similar $M(H)$ dependences were found for rolled tantalum [21]. The observed form of the magnetization curve can be also resulted due to simultaneous regions with type-I and II superconductivity in the sample. To clarify, we compare the field dependences of irreversible $M_{irr}$ and reversible $M_{eq}$ magnetizations of the samples S1 and S2 (Fig. 4). The $M_{irr}$ and $M_{eq}$ values are defined as $M_{irr} = (M_\downarrow(H) - M_\uparrow(H))/2$ and $M_{eq} = (M_\downarrow(H) + M_\uparrow(H))/2$, where $M_\uparrow(H)$ is the forward branch and $M_\downarrow(H)$ is the reversal branch of the magnetization hysteresis loop.

The both samples have the similar $M_{irr}$ contribution at $0 < H < H_c$ that corresponds to the magnetic flux trapping by superconducting contours [20]. The $M_{irr}(H)$ dependence for S2 also has a smooth region at $H_c < H < H_{c2}$ that is due to flux pinning into the particles with type-II superconductivity. The $M_{eq}(H)$ curve of S1 is qualitatively similar to the initial section of the $M_{eq}(H)$ curve of S2 (at $\mu_0 H < 0.02$ T). Both curves achieve minimum values in similar way. Correlated behavior of the $M_{irr}(H)$ and $M_{eq}(H)$ curves at low $H$ region confirms that the nanoparticles with type-I superconductivity are contained in S2.

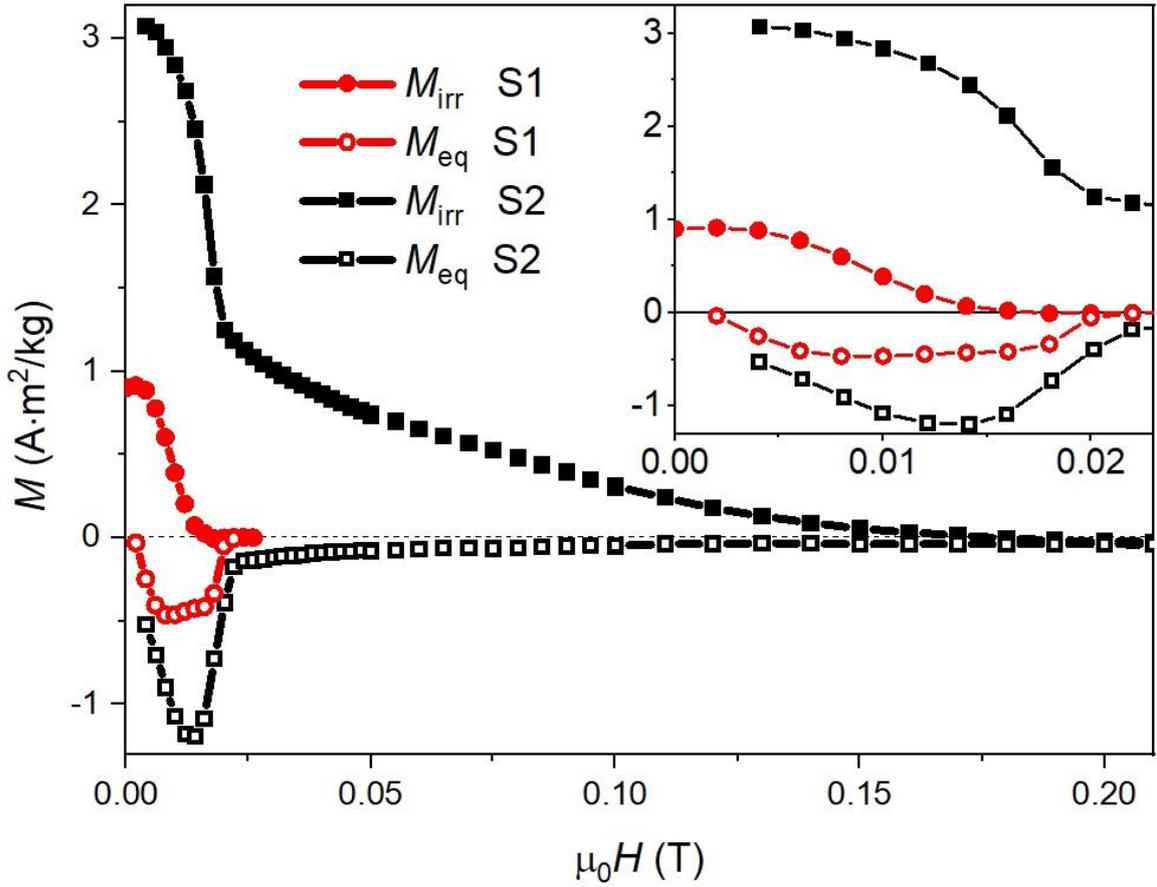

Fig. 4. The field dependences of the irreversible $M_{irr}$ and the reversible $M_{eq}$ magnetization. The insert shows the same curves in small fields.

With $\lambda_L = 35$ nm as for the bulk tin [1] and the size > 50-70 nm, nanoparticles are the type-I superconductor with $\kappa < 1/\sqrt{2}$. There are only particles larger than 50 nm in the sample S1. The sample S2 contains nanoparticles both larger and smaller than 50 nm. So the regions with the type-I and II superconductivity are supposed to coexist in S2. This leads to the analog of the intertype superconductivity for magnetic properties of the sample S2.

Generally, materials with magnetic properties corresponding to an intertype superconductor can be realized as a mixture of type-I and II superconducting particles. A form of the magnetization curve is controlled by the distribution of the particle size. The ratio of the smaller and larger nanoparticles can affect the different sections of the magnetization curve.

**Conclusions**

In conclusion, superconducting IOS samples were investigated. The sample S1 with the 128 nm and 70 nm tin nanoparticles is the type-I superconductor. The magnetic properties of the sample S2, which contains the 80 nm and 42 nm tin nanoparticles with type-I and II superconductivity correspondingly, are similar to characteristics of intertype superconductors.

The size distribution of superconducting nanoparticles should be accounted to interpret intertype-like magnetization curves [21].

**Acknowledgements**

The authors are grateful to D.A. Balaev, N.E. Savitskaya, N.A. Grigorieva for fruitful discussions and to A.A. Mistonov for provided samples. The work is supported by the Russian Science Foundation (project No. 17-72-10067).